\RequirePackage{fix-cm}
\documentclass[twocolumn,epjc3]{svjour3}  
\smartqed  % flush right qed marks, e.g. at end of proof
\RequirePackage{graphicx}
\usepackage{subfig}
\usepackage{amsfonts}
\usepackage{tikz}

\RequirePackage[colorlinks,citecolor=blue,urlcolor=blue,linkcolor=blue]{hyperref}
\RequirePackage{lineno}
\usepackage{amsmath}
\journalname{Eur. Phys. J. C}

\begin{document}
%\setpagewiselinenumbers

%\linenumbers
\title{A novel approach for nearly-coincident events rejection}


\author{
	M.~Borghesi \thanksref{e1,addr2,addr3} \and
	M.~De~Gerone \thanksref{addr1} \and 
	M.~Faverzani \thanksref{addr2,addr3} \and
	M.~Fedkevych \thanksref{addr1} \and
	E.~Ferri    \thanksref{addr2,addr3} \and 
	G.~Gallucci \thanksref{addr1} \and
	A.~Giachero \thanksref{addr2,addr3} \and
	A.~Nucciotti \thanksref{addr2,addr3} \and
	%        G.~Pessina  \thanksref{addr4} \and
	A.~Puiu     \thanksref{addr4,addr5}
}

\thankstext{e1}{e-mail: matteo.borghesi@mib.infn.it}


\institute{Istituto Nazionale di Fisica Nucleare (INFN), Sezione di Genova, Genoa 16146, Italy \label{addr1} \and
Dipartimento di Fisica "G. Occhialini", Universit\`{a} di Milano - Bicocca, Milan 20126, Italy \label{addr2} \and
Istituto Nazionale di Fisica Nucleare (INFN), Sezione di Milano-Bicocca, Milan 20126, Italy \label{addr3} \and
Gran Sasso Science Institute (GSSI), I-67100 L'Aquila, Italy;
\label{addr4} \and 
INFN - Laboratori Nazionali del Gran Sasso, Assergi (L'Aquila) I-67010 - Italy
\label{addr5}
}

\date{Received: date / Accepted: date}

\maketitle

\begin{abstract}
We present a novel technique, called DSVP (Discrimination through Singular Vectors Projections), to discriminate spurious events within a dataset. The purpose of this paper is to lay down a general procedure which can be tailored for a broad variety of applications. After describing the general concept, we apply the algorithm to the problem of identifying nearly coincident events in low temperature microcalorimeters in order to push the time resolution close to its intrinsic limit. In fact, from simulated datasets it was possible to achieve an effective time resolution even shorter than the sampling time of the system considered. The obtained results are contextualized in the framework of the HOLMES experiment, which aims at directly measuring the neutrino mass with the calorimetric approach, allowing to significally improve its statistical sensitivity.

%The HOLMES experiment is a large-scale experiment for the electron neutrino mass determination. It will perform a calorimetric measurement of the energy released in the electron capture decay of 163Ho. In its final stage, HOLMES will employ 1000 microcalorimeters with Transition Edge Sensors (TES). These detectors are being used more and more frequently in physics and astronomy experiments, due to their energy resolution and their multiplexing capability. However, their excellent intrinsic energy resolution cannot be preserved without  an accurate analysis procedure. 
%Each of the HOLMES detector will be implanted with an activity of 300 Hz. The events will be recorded with a sampling frequency of 500 kHz, corresponding to 1024 points acquired in 200 microseconds. The purpose of signal processing is to extract as many information as possible from those events.
%This contribution will provide an overview of our algorithms used for pulse processing, from the evaluation of pulses energy to pile-up rejection. With the HOLMES high decay rates, reliable identification of nearly-coincident events is crucial to suppress what is expected to be the leading source of background and systematic errors. We report here the time resolution obtained with the combination of Wiener Filter and a processing method that exploits singular value decomposition.
\end{abstract}

\section{Introduction}
\label{sec:1}

The demand for increasing sensitivities of nowadays experiments requires the development of complex analysis tools to respond to several demands, according to the design and goals of the experiment. In many experiments, a crucial factor in achieving a high sensitivity is the ability to discriminate spurious events. This is a particularly relevant feature to keep into account for experiments where the statistics of the spurious events might be comparable with, or even overcome, the statistics of the proper events. This is the case, for instance, of the direct measurement of the neutrino mass with the calorimetric approach \cite{nucciotti2016use}.

\noindent So far, few techniques are currently employed for the purpose of discriminating the spurious events from the proper ones, and they all require that events belonging to these two families must differ in some way from each other.
In this paper we outline a novel technique, called DSVP (Discrimination through Singular Vectors Projections), based on a previous work by Alpert et al. \cite{Alpert:2015mtu}.
Besides its effectiveness, one of the main advantage of this method, is that it does not rely on any particular assumption about the structure of the events, thus in principle it can be applied in various scenarios almost in a semi-automatic way.

\noindent In section \ref{sec:2} we  illustrate the DSVP method while in
%in a general manner, going through the algorithm step by step. In 
the second part of the article (Section \ref{sec:3} and \ref{sec:4}) we show an application of this technique in the context of a direct calorimetric neutrino mass measurements. In particular, we will focus  on the simulated data sets representing the one foreseen for detectors used in the HOLMES experiment \cite{alpert2015holmes}, for which the expected main source of background will be unrecognized pile-up events. 

\section{The DSVP technique}
\label{sec:2}

\noindent The aim of the DSVP algorithm is to discriminate as many as possible undesirable events (i.e. spurious events which differ from a reference signal) present in a given dataset, using the information about the mean 'morphology' of the events present in the dataset.

\noindent In order to apply the DSVP technique, the following elements are required:
\begin{itemize}
	\item The measured  dataset, $\vec{M}$. This $n \times d$ matrix consists of the  dataset of interest, where each row is an event described by $d$ variables. Namely, the events can be seen as points in the  $\mathbb{R}^d$ space. From now on, we call the \textit{good} events in the dataset $A$ events, while $B$ events are the ones to be rejected. 
	We assume that the $A$ events are more numerous respect to the $B$ ones ($N_A > N_B$) \footnote{See section \ref{sec:2a}}.
	
	\item The expected number of $B$ events $N_B$ that the algorithm should discard at most.  
	\item A training dataset, $\vec{T}$, such that $N_A >> N_B$. The events of this $n'\times d$ matrix can be distributed in a different region of $\mathbb{R}^d$ respect to the events in $\vec{M}$. For instance, in the case of microcalorimeter signals, the events in $\vec{T}$ can lie in a different energy range respect to the events in $\vec{M}$.
\end{itemize}

\noindent We will use the training dataset $\vec{T}$ to define a new vector space which will help us to highlight the features that distinguish an $A$ event from a $B$ one.
\noindent This new vector space, called from now the projections space, has dimension $k$, with $k << d$. The events can be represented as points in the  $\mathbb{R}^k$ projection space, so that the $A$ events are distributed differently respect to the $B$ ones.
The idea is to find a model (i.e. hypersurfaces) describing the distribution of the $A$ points in $\vec{M}$ in this new space, so that the $B$ points can be identified as the ones with a larger distance from what predicted by the model.

%\noindent In order to find the parameters defining the model first we need to 'clean' the dataset,
\noindent  In order to find the model we need to 'clean' the dataset first, obtaining a subset of $\vec{M}$, $\vec{M'} \subset \vec{M}$, which contains mostly $A$ events at the expense of deleting also some $A$ events. %This process is described in detail in section \ref{sec:2a}.

\noindent The next step is to represent the events in $\vec{M'}$ in the projection space and to find the model parameters which describes the distribution of the $\vec{M'}$ ($\sim A$) events. 

\noindent We then define the discrimination parameter and its threshold to recognize an $A$ event from a $B$ one in $\mathbb{R}^k$.

\noindent Finally, we take the original dataset $\vec{M}$, represent the events in the projection space, find the discrimination parameter and discard all the events that have a value of the discrimination parameter above the threshold found.

\noindent The procedure (dataset 'cleaning', model and threshold definition and $B$ discrimination) is then repeated with the survived events. At each iteration, the $\vec{M}$ dataset will contain a smaller fraction of $B$ events.

\noindent In the following sections, each steps of the algorithm are described in detail.

\subsection{Raw cleaning with PCA}
\label{sec:2a}
%This matrix will be used to define the model which describes the $A$ events in the next section.
In the first step, the aim is to create a suitable dataset for modeling the distribution of $A$ events in the $\vec{M}$ matrix, lowering the ratio $N_B/N_A$ at the expense of deleting also $A$ events.
Knowing that $N_A > N_B$,  the mean 'morphology' of the events is closer to the $A$ ones. We can define a suitable parameter using the Principal Component Analysis (PCA) \cite{Jolliffe2011} to discard mainly $B$ events.

\noindent The procedure used is equal to the one described in \cite{Alpert:2015mtu}, which will be reported for completeness.
The singular value decomposition (SVD) \cite{10.5555/248979} is computed for the $n \times d$ matrix $\vec{M}$, which is decomposed in a product of three matrices $\vec{M}=\vec{U}\vec{D}\vec{V^T}$. The columns of $\vec{U}$ and $\vec{V}$ are the left and right singular vectors respectively, while the entries of the diagonal matrix $\vec{D}$ are the singular values. The singular values are ordered from 1 to $d$ in order of importance. Only the first $j<d$ columns of $\vec{D}$ are non-neglibile. It is convenient to define a new matrix  $\hat{\vec{U}}$ which contains only the first $j$ columns of $\vec{U}$ subtracted by their means which is equivalent to centering the data matrix, as required by the PCA.

%\begin{figure}[h!]
%	\centering
%	\includegraphics[width=1\linewidth]{Singular_values_distribution.pdf}
%	\caption{An example of the decreasing trend of the singular values of the matrix $\vec{M}$}
%	\label{fig:singularvalues}
%\end{figure}

\noindent The columns of $\hat{\vec{U}}$ are vectors of length $n$. Basically, they represent the projections of the mean-centered events contained in $\vec{M}$ on the right singular vectors (i.e. the columns of $\vec{V}$, which are called principal vectors in the PCA framework) with the first column of $\hat{\vec{U}}$ expressing the projections on the first right singular vector and so on. The columns of $\vec{V}$ are vectors of dimension $d$ which represent the direction of greatest variance of the data in $\vec{M}$.
Thanks to the properties of the PCA, an appropriate combination of the projections can be of use to define a parameter, called $norm^2$, which indicates how close an event is to the mean 'morphology' of the events in $\vec{M}$.  

\noindent The precision matrix $(\sigma^2)^{-1}$ is computed from the $j \times j$ empirical covariance $ \sigma^2 = \hat{\vec{U}}^T \hat{\vec{U} }$ and it is used to evaluate the parameter $norm^2$ for each event $i = 1,...,n$ in the matrix $\vec{M}$
\begin{equation}
	{norm}^2_i = \hat{\vec{U}}_{i,*} (\sigma^2)^{-1}  \hat{\vec{U}}^T_{i,*} 
\end{equation}
Suppose that we have a guess of how many $B$ events are expected in the dataset. We call this number $N_{B}^{guess}$. 
$B$ events deviate disproportionately from the mean in this covariance-adjusted sense, so we discard those with largest $norm^2$ and repeat the procedure on the remaining data a total of $l$ times, removing on the $l$-th iteration a number of events equal to $N_{B}^{guess}/2^{l}$ with the largest $norm^2$. In our tests, we use $l=5$.

\noindent The iterations guarantee that the mean morphology of the events are closer and closer to the $A$ events each cycle, as $B$ events are increasingly eliminated.

\noindent After the PCA, we have eliminated
\begin{equation}
	N_{del}^{PCA} = \sum_l \frac{N_{B}^{guess}}{2^l} 
\end{equation}
events, where the $B$ events are the ones predominantly discarded. The remaining events after the PCA are $m= n - N_{del}^{PCA}$. We call $\vec{M'}$ the $m \times d$ the matrix of the survived events, which is mostly composed of $A$ events. %and in the next section it will be used to define a model describing them.text

%In the next section, $\vec{M'}$ in combination with the training matrix $\vec{T}$ will be used to define a model which describes the $A$-type events. With this model, the $A$-type events eliminated by the PCA will be recovered.

\subsection{Define a model for the A-events} 
\label{sec:2b}
To discriminate the undesirable events, we now need to define a model which describes the distribution of the $A$ points (the ones belonging to $\vec{M'}$) in the projection space.

\noindent First, we need to define this space. We decompose the $\vec{T}$ matrix using the SVD. Because the training matrix $\vec{T}$ is mainly composed by $A$ events, we assume that its first $k$ significant right singular vectors $\{ \vec{v}_1,\vec{v}_2,...,\vec{v}_k \}$ can constitute a base of the projection space.

 %we assume that the vectors which describe its data, the right singular vectors of $\vec{T}$, will help us to recognize the proper signals in $\vec{M}$. Therefore, the vectors that constitute a base of the projection space are obtained by applying the SVD to $\vec{T}$ and are the first $k$ significant right singulars vectors $\{ \vec{v}_1,\vec{v}_2,...,\vec{v}_k \}$.

\noindent The events in $\vec{M'}$ are projected onto these vectors. 
From now on, each event in  $\vec{M'}$ will be described by $k < d$ variables, its projections onto the right singular vector of $\vec{T}$. We indicate all the coordinates of the $\vec{M'}$ points along the $i$-th base vector of the projection space as  $p_i = \vec{M'} \cdot \vec{v_i}$.

%Because the right singular vectors come from a matrix basically made of $A$-type events, the new $k$ data describing the $A$-type events are distributed differentely compared to the $B$-type ones.

\noindent %Describing the points distribution in a high dimensional space is not trivial. We decided to do it by dividing the projection space $\mathbb{R}^k$ in different subspace and describing the distribution of the $\vec{M'}$ points with a set of curves, one for each subspace.
\noindent To describe the points distribution in the new vector space, the projections $p$ are classified into two groups: the $k'$ independent projections, indicated as $\vec{p}_{ind}$ and the dependent ones, $\vec{p}_{dep}$.

\begin{equation}
\{p\}_k = \underbrace{p_1 , ... , p_{k'}}_{\vec{p}_{ind}} ,\underbrace{p_{k'+1} , ...,p_k}_{\vec{p}_{dep}}
\end{equation} 

\begin{figure}[h!]
	\centering
	\includegraphics[width=1\linewidth]{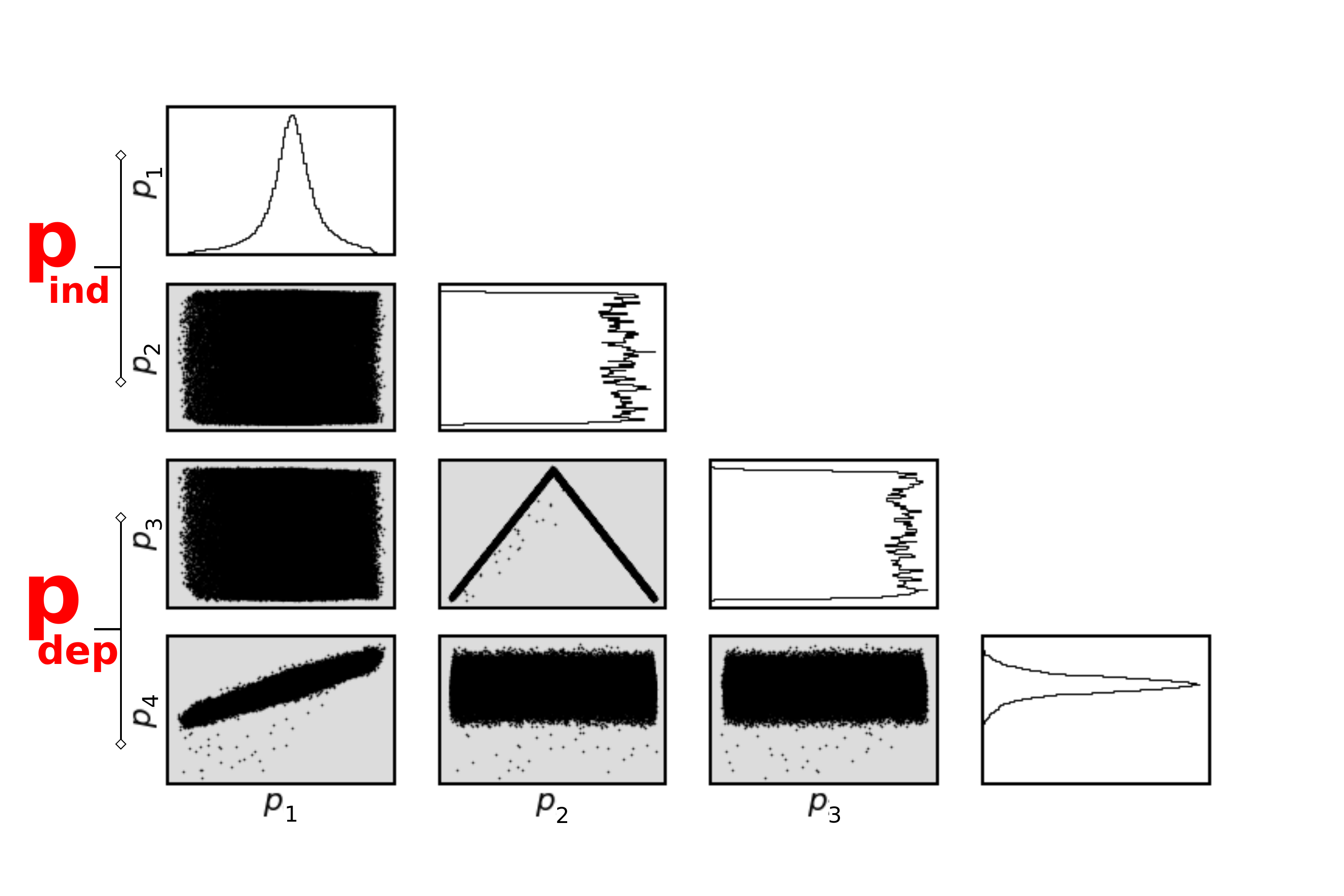}
	\caption{An example of distribution of the points in $\vec{T}$ in the projection space from sec. \ref{sec:3}. In this particular case, the projection space is in $\mathbb{R}^4$. We decided to set $k' =2$, thus describing the points distribution in $\mathbb{R}^4$ with two curves: $p_3 =f_3(p_1,p_2)$ and $p_4 = f_4(p_1,p_2)$}
	\label{fig:projdistT}
\end{figure}

\noindent The dependent projections can be expressed as a function of the independent ones. There is no general rule to identify which projection is "independent" and which one is not, since it is related to the specific problem. The training dataset can be used to identify the dependencies among the projections, as shown in Fig \ref*{fig:projdistT}.

\noindent The distribution of the dependent projections can now be easily described in a $\mathbb{R}^{k'+1}$ subspace by a set of $f$ curves

\begin{equation}
p_{i} = f_i(\vec{p}_{ind}) \ \ \ ; \ \ \  i=k'+1,...,k
\end{equation}
Knowing precisely the set of curves $\{f\}$, we will be able to differentiate between the two distribution of events, because the projection of the $B$ events will not follow the same curves as the one of the $A$ events.

\noindent Usually the functional form of the different $f$ is unknown. However, we can approximate each $f$ curve with a Taylor expansion and let a (weighted) linear regression find the best parameters of the expansion. In particular, we use a modified version of the random sample consensus (RANSAC) algorithm \cite{Ransac} \footnote{Due to the fact that the number of outliers ($B$ events) from step \ref{sec:2a} is expected to be negligible, any type of weighted linear regression can in principle be used.}.

\noindent The set of curves $\{f\}$ which describes the $\vec{M'}$ events in the projection space is what we called the model. 
%Once the model is evaluated, we can recover the $N_{del}^{I,A}$ events.

\subsection{Find a discrimination threshold}
\label{sec:2c}
The difference between the measured dependent projections and the ones expected from the model is evaluated for each event in the $\vec{M'}$ matrix. A residual norm is defined as
\begin{equation}
d = \sqrt{\sum_{j=k'+1}^{k} (p_{j}-f_j(\vec{p}_{ind}))^2} 
\end{equation}
In order to discriminate between the $A$ events, the one with the lowest residual norm, and the $B$, we need to define a threshold value, $d_{thr}$. 
Due to the fact that the $\vec{M'}$ dataset is mainly made of $A$ events the threshold is chosen as the highest value of $d$ plus the standard deviation of the $d$ distribution

\begin{equation}
\label{thr_def}
d_{thr} = \textrm{max} \{d\} + \textrm{std}\{d\}
\end{equation}
This threshold definition should ensure to include not only the $A$ events in $\vec{M'}$, but also the $A$ events in the original dataset $\vec{M}$ which were eliminated by the 'PCA cleaning' described in \ref{sec:2a}. Nevertheless, this definition of threshold might need to be redefined to account for the specific problem considered.

\subsection{Apply the model}
\label{sec:2d}
\noindent Now all the components to make the algorithm work are present: a base for the projection space, a set of curves to model the points distribution in that space and a discrimination threshold.

\begin{figure*}[h!]
	\subfloat[\label{fig:sfig1}]{%
		\includegraphics[width=.46\linewidth]{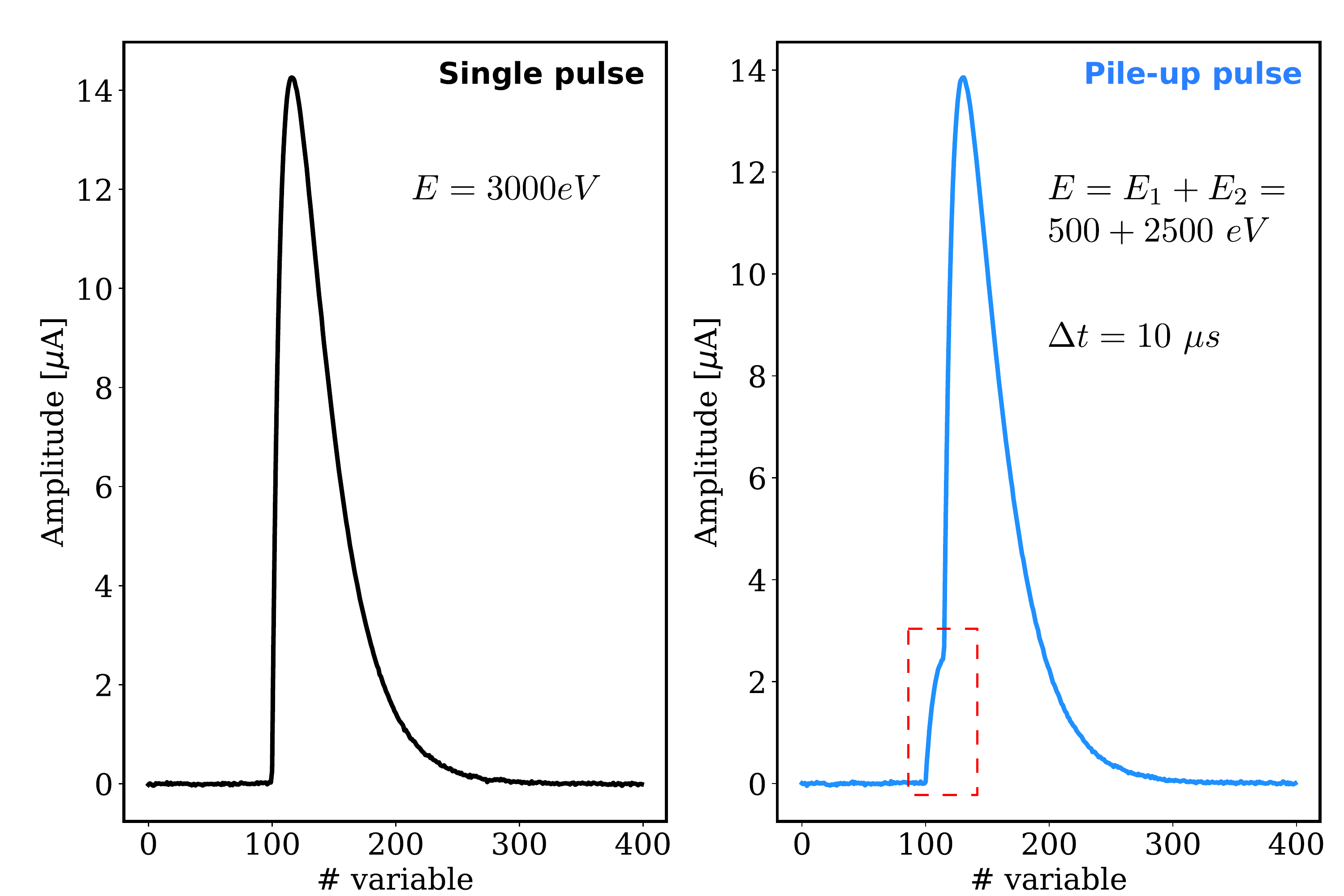}%
	}\hfill
	\subfloat[\label{fig:sfig2}]{%
		\includegraphics[width=.46\linewidth]{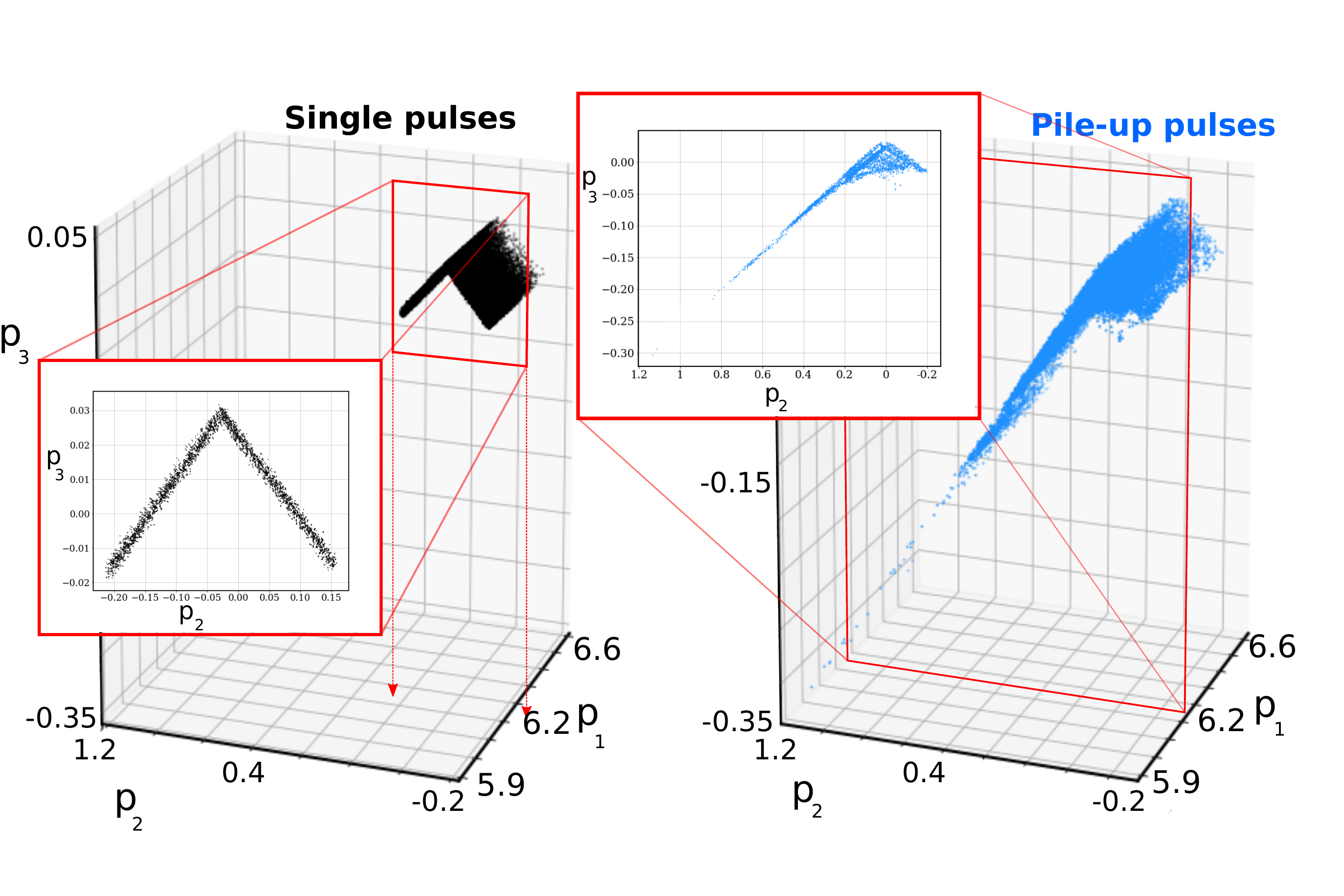}%
	}
	
	\subfloat[\label{fig:sfig1}]{%
		\includegraphics[width=.46\linewidth]{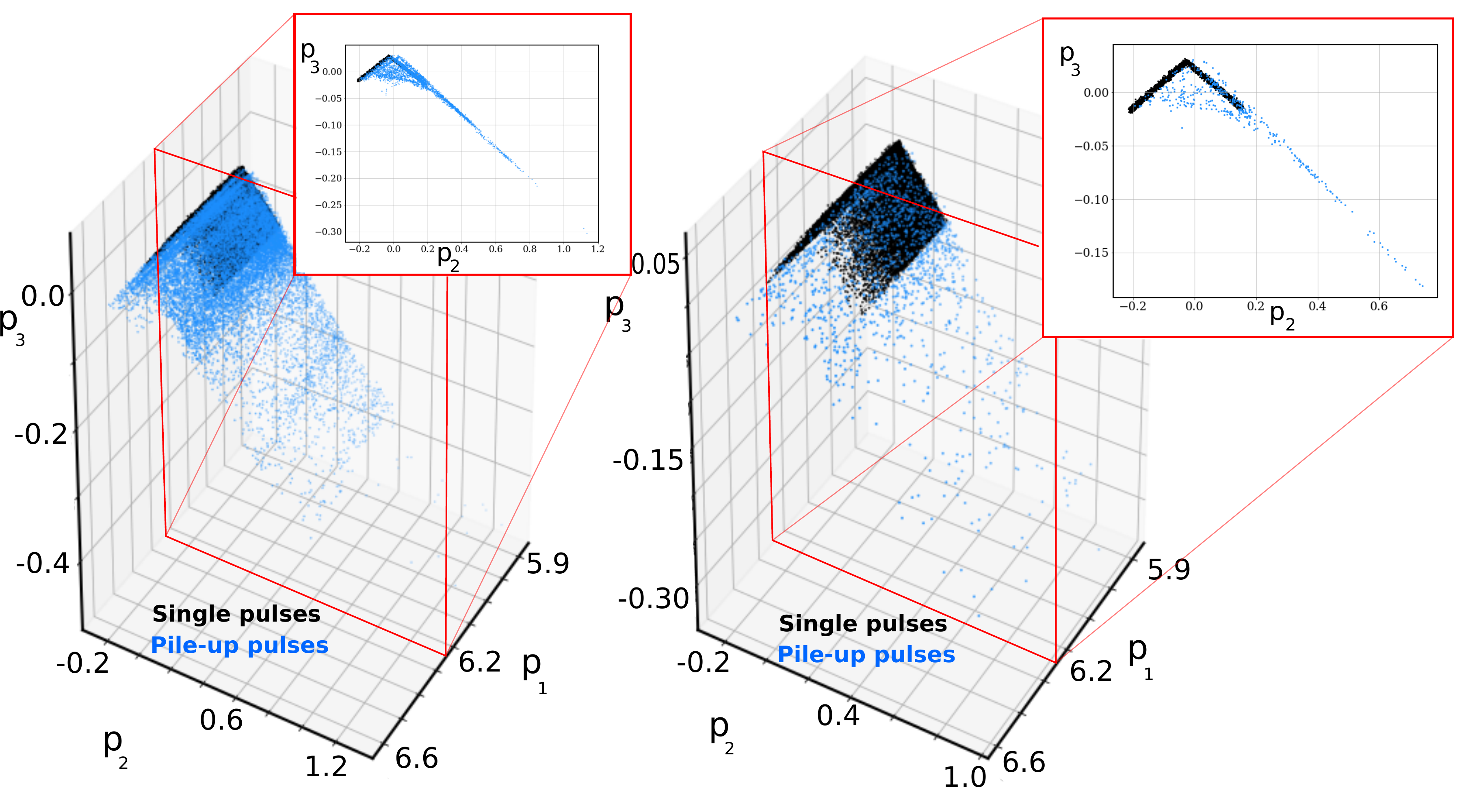}%
	}\hfill
	\subfloat[\label{fig:sfig2}]{%
		\includegraphics[width=.46\linewidth]{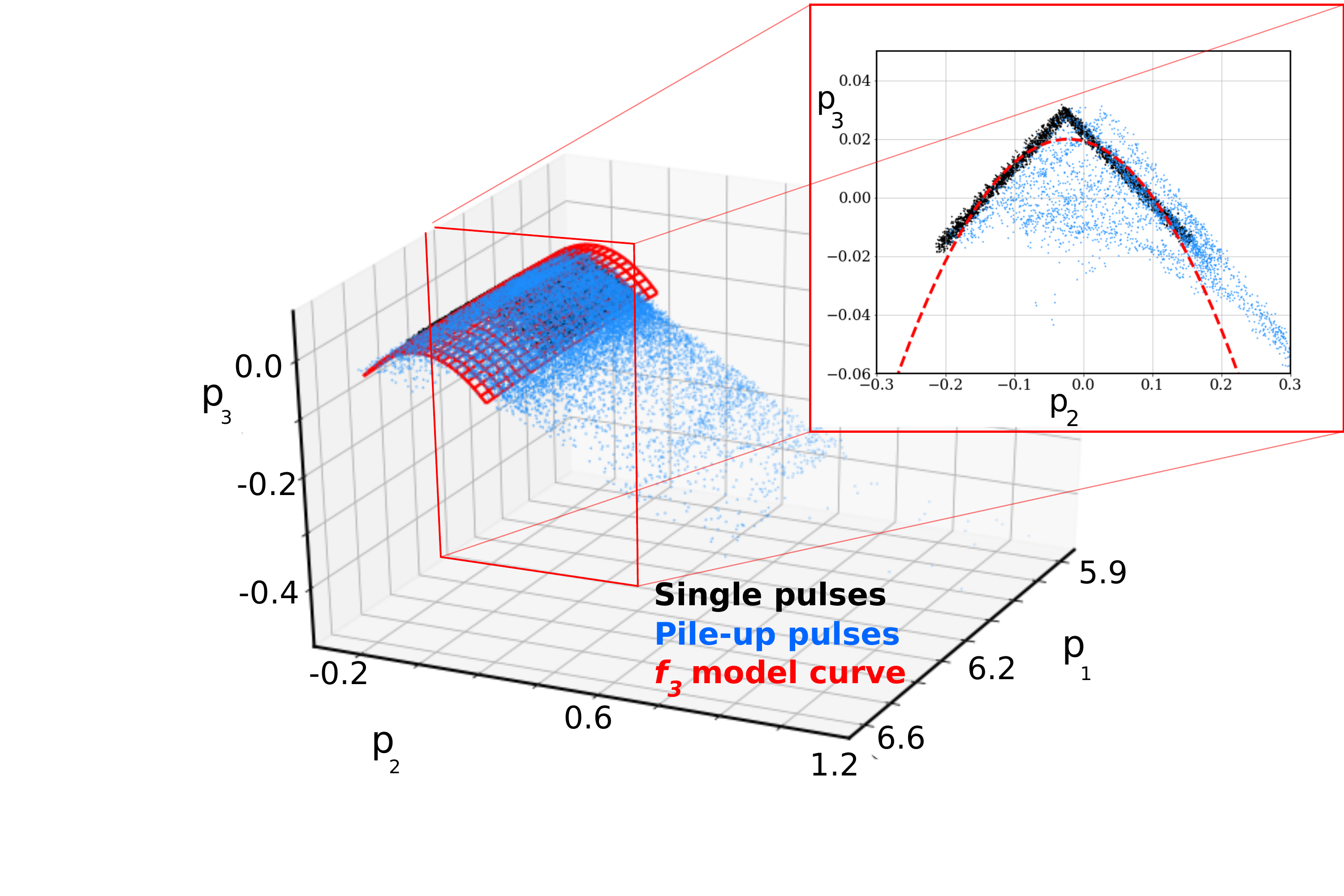}%
	}
	\caption{a)  Each event (row) of the matrix $\vec{M}$ is initially described by 400 variables i.e. samples; we can represent each event as shown in figure or as points in $\mathbb{R}^{400}$. $\vec{M}$ contains two different types of events: single pulses ($A$) of energy $E$ and pile-up pulses with different arrival time ($B$) with energies $E_1$ and $E_2$ such that $E_1+E_2=E$. (b) The events of $\vec{M}$ are represented in the projections space. In this space, the two types of events follow two different distributions. (c) In the left (right) panel the matrix $\vec{M}$ ($\vec{M'}$) is represented in the projection space. It is possible to appreciate how the PCA has drastically reduced the fraction of pile-up.	(d) The curve $f_3 = f_3 (p_1,p_2)$, which describes the distribution of the events in $\vec{M'}$, is used to discriminate between the single pulse and pile-up pulses.}
	\label{fig:2}       % Give a unique label
\end{figure*}

\noindent We will now use these tools on the original dataset $\vec{M}$, namely:
\begin{enumerate}
	\item Take the inner product of the events in $\vec{M}$ with the base of the projection space, determining $p_k$.
	\item Evaluate the residual norm $d$ using the curves describing the $A$ projections distributions.
	\item The events with a residual norm above the threshold are discarded.
\end{enumerate}
After the third step, we will have discarded $N_{del}$ events.
% with $N_{del} \leq N_{del}^{PCA}$. %Given the threshold definition \eqref{thr_def},
The events deleted by the third step will be almost, if not all, spurious $B$ events.
All the previous steps (PCA, model and threshold definition) are now repeated with a reduced number of expected $B$ events, $N^{B'} = N^B - N_{del}$. The iterations successively improve the representation of $A$ events, as $B$ events are increasingly eliminated.
The algorithm stops when $N_{del} = 0$ or when $N^{B'} = 0$.

\noindent Fig \ref{fig:2} shows a visual representation of some of the steps of the DSVP technique. As an example the figure reports signals from a TES microcalorimeter, as explained in \ref{sec:3b}. This particular case was chosen because there are just 3 non-negligible singular values, therefore the points in the projection space can be easily shown on a 3D plot.

\noindent The method was implemented in python, taking advantage of many of the fast modules of NumPy and SciPy. The majority of the computational time is taken by the \textit{Raw cleaning with PCA} part, due to the fact that the SVD on the matrix $\vec{M}$ is performed five times for each iteration. Nevertheless, the algorithm is quite fast, taking $\sim$ 7 minutes to compute 9 iterations on a matrix $\vec{M}$ composed of 120000 rows and 1024 columns of float32 numbers, using only one (six years old) CPU with a base clock of 2.6 GHz.

\section{HOLMES and pile-up discrimination}
\label{sec:3}
The algorithm described in sec \ref{sec:2} is now applied in the framework of HOLMES. The HOLMES experiment will perform a direct measurement of the neutrino mass with a sensitivity of the order of 1 eV,  measuring the energy released in the electron capture (EC) of $^{163}$Ho, as proposed by De Rujula and Lusignoli in \cite{de1982calorimetric}.
It will also demonstrate the scalability of the calorimetric technique for a next generation experiments that could go beyond the current best expected sensitivity of 0.1 eV \cite{aker2019improved}.
In order to reach this sensitivity, HOLMES will use low temperature TES microlorimeters with $^{163}$Ho implanted in their absorbers, with an activity of 300 Hz per detector. 

\noindent The effect of a non-zero neutrino mass on the $^{163}$Ho EC decay spectrum can be appreciated only in a energy region very close to the end point, where the count rate is low and the fraction of nearly-coincident events, called pile-up events, to single events is greater than one. 
\noindent If a pile-up event is composed of two events of energies $E_1$ and $E_2$ which occur within a time interval shorter than the time resolution of the detector, it is recorded as a single event with energy $E \simeq E_1 + E_2$. Thus, if not correctly identified, pile-up events will distort the decay spectrum of $^{163}$Ho, lowering the sensitivity to $m_{\nu}$. 

\begin{figure}[h!]
	\centering
	\includegraphics[width=1.\linewidth]{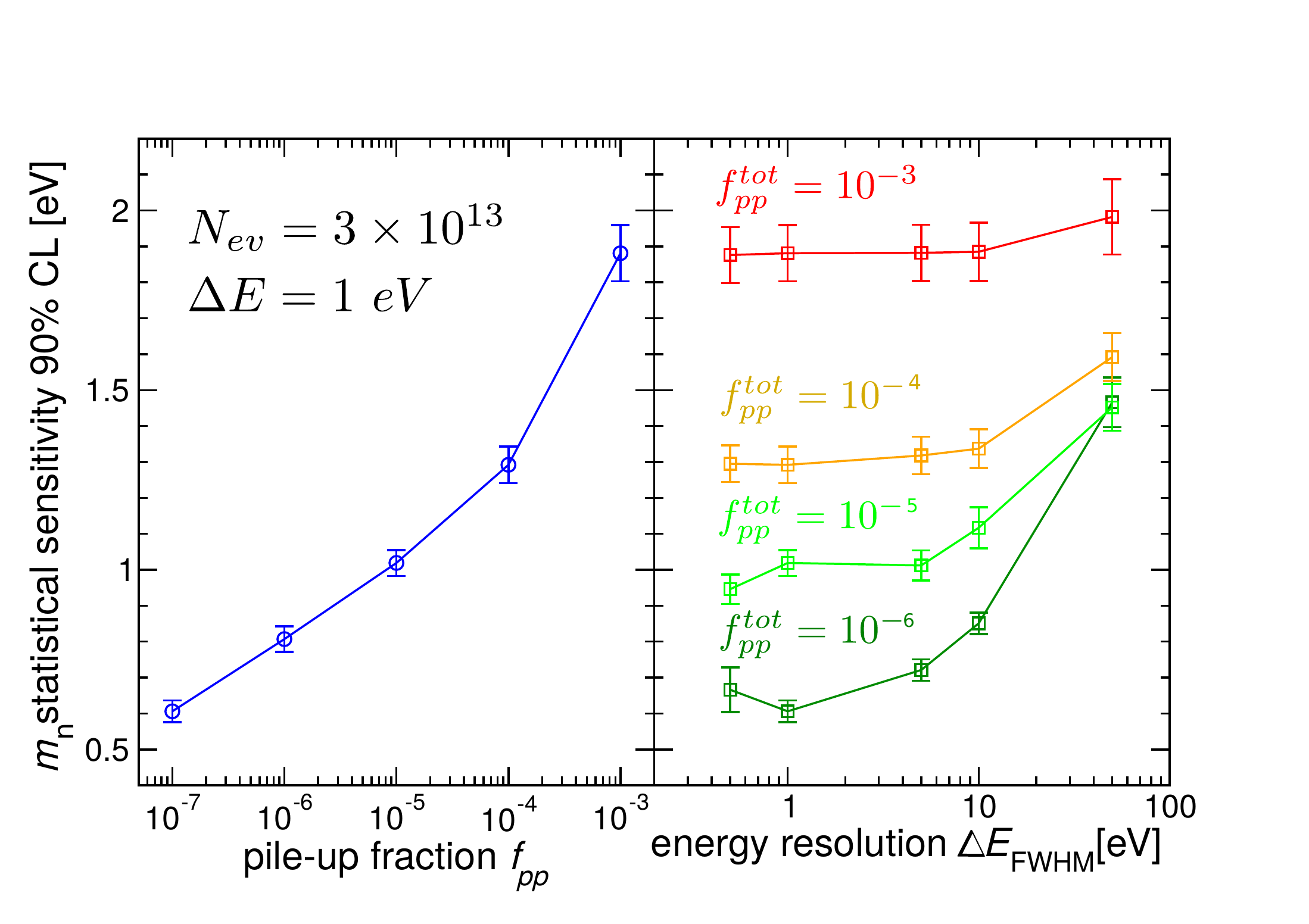}
	\caption{Expected neutrino mass sentivity for the HOLMES experiment. The right panel shows the sensitivity compared to the detectors energy resolution at different pile-up fraction. We called the pile-up fraction  on the whole energy spectrum $f_{pp}^{tot}$. }
	%Red $f_{pp} = 10^{-3}$, yellow $f_{pp} = 10^{-4}$, light green $f_{pp} = 10^{-5}$ and dark green $f_{pp} = 10^{-6}$.}
	\label{fig:sensitivity}
\end{figure}

\noindent The neutrino mass sensitivity of HOLMES has been evaluated through Monte Carlo simulations \cite{nucciotti2014statistical}, see Fig \ref{fig:sensitivity}. Once fixed the number of recorded events to $3\times 10^{13}$, the simulations have shown that the sensitivity on neutrino mass is not strongly dependent on the energy resolution of the detector (as long as $\Delta E < 10$ eV), but rather on the pile-up fraction $f_{pp}$, i.e. the ratio between the number of pile-up events to single events. Its reduction is crucial for the success of the experiment.

\noindent Using the terminology of section \ref{sec:2}, in the HOLMES experiment an $A$ event is a signal caused by a single energy deposition in the microcalorimeter detector, while a $B$ event is a signal caused by nearly coincident events. Each signal is a collection of  records $I_i$ of the current flowing through the detector sampled at an instant $t_i = i \times t_{samp}$, where $t_{samp}$ is the sampling time of the readout system. An example of a microcalorimeter signal is shown in Fig \ref{fig:2} (a). With the current setup, the sampling time is fixed at 2 $\mu$s. 

\noindent We tested the algorithm robustness and efficiency through many simulations which aim at emulating the results expected by the HOLMES experiment. The HOLMES TES microcalorimeters do not have the $^{163}$Ho implanted yet, therefore a real data test will be done at later times.

\subsection{Energy spectrum \& ROI definitions}
\label{sec:3a}

$^{163}$Ho decays via electron capture to an atomic excited state of $^{163}$Dy which relaxes mostly emitting atomic electrons (i.e. the fluorescence yield is less than 10$^{-3}$ \cite{de1982calorimetric}). The de-excitation energy $E_{c}$ spectrum probability density is proportional to 
\begin{equation}
\frac{d\lambda_{EC}}{dE_c} \propto \sqrt{(Q-E_c)^2-m_\nu^2}
\end{equation}
where $m_{\nu}$ is the effective neutrino mass.
\noindent The pulses are generated according to the spectrum in \cite{de1982calorimetric} with $Q=2.833$ keV \cite{eliseev2015direct}, $m_\nu=0$ and with energy between 2.650 keV and 2.900 keV. Second order effects like shake-up and shake-off \cite{gastaldo2017electron} have not been considered in this work.
Despite the optimal region of interest (ROI) aimed at determining the neutrino mass will be determined only when actual data will be collected, this energy range can be considered as a reasonable ROI.

\noindent Each detector must be treated separately with the DSVP technique in order to account for their slightly different characteristics. Thus, to create a statistic expected for a single detector with a target activity of 300 Hz over two years of data taking, we generated 40000 ($\sim$ 80000) single (double) pulse events. The arrival time of the pile-up pulses is uniformly distributed between 0 and 10 $\mu$s.

\subsection{Detector models}
\label{sec:3b}

\noindent For this study we modeled three different TES microcalorimeters with the one-body model \cite{irwin2005transition} or with the two-body dangling model \cite{maasilta2012complex}. %depending on whether an unknown parasite heat capacity is present or not.
%the finite thermal conductance between the absorber and the thin film TES is negligible or not.
In both cases the current pulse profile is obtained by solving the system of the electro-thermal differential equations applying the fourth-order Runge-Kutta method (RK4) and considering the transition resistance as proposed by \cite{cabrera2008introduction} for taking into account the TES non-linear behavior. To these pulses a noise waveform, generated as an autoregressive moving average ARMA (p,q) process %on Gaussian white noise and 
with a power spectrum given by the Irwin-Hilton model, is added. 

\begin{figure}[h!]
	\subfloat[\label{fig:sfig2}]{%
	\includegraphics[width=1\linewidth]{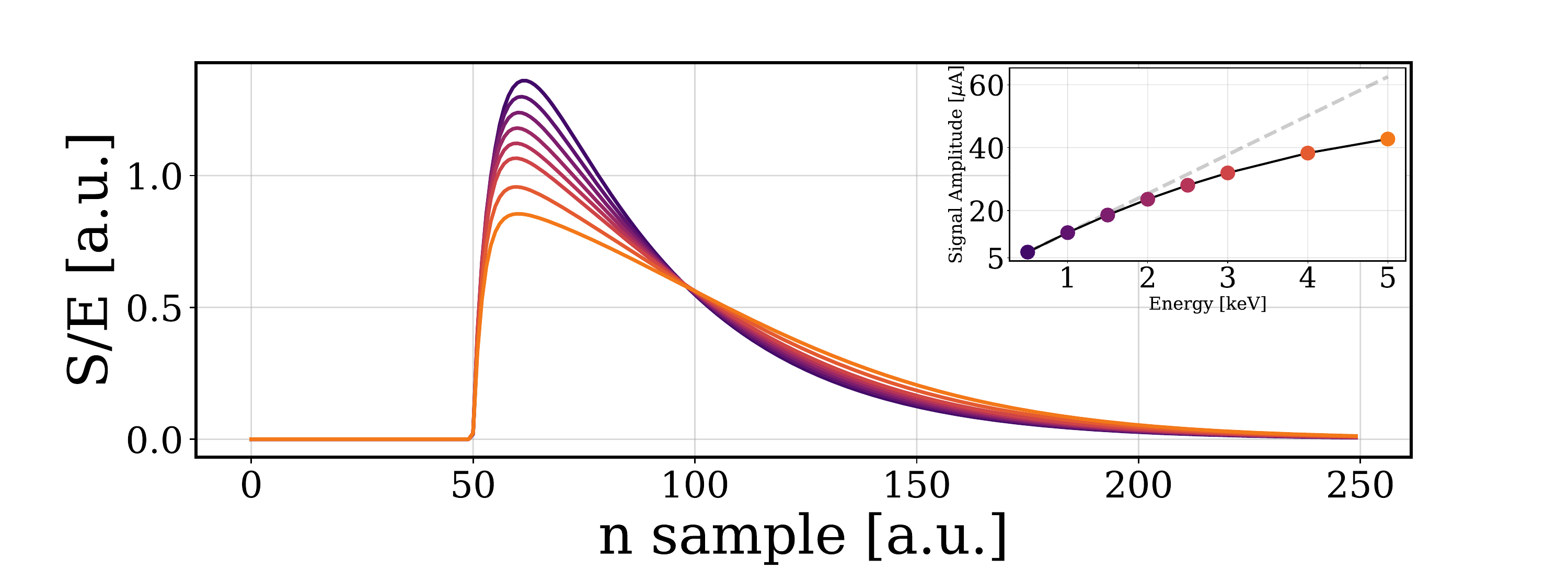}%
}
\vfill
	\subfloat[\label{fig:sfig1}]{%
	\includegraphics[width=1\linewidth]{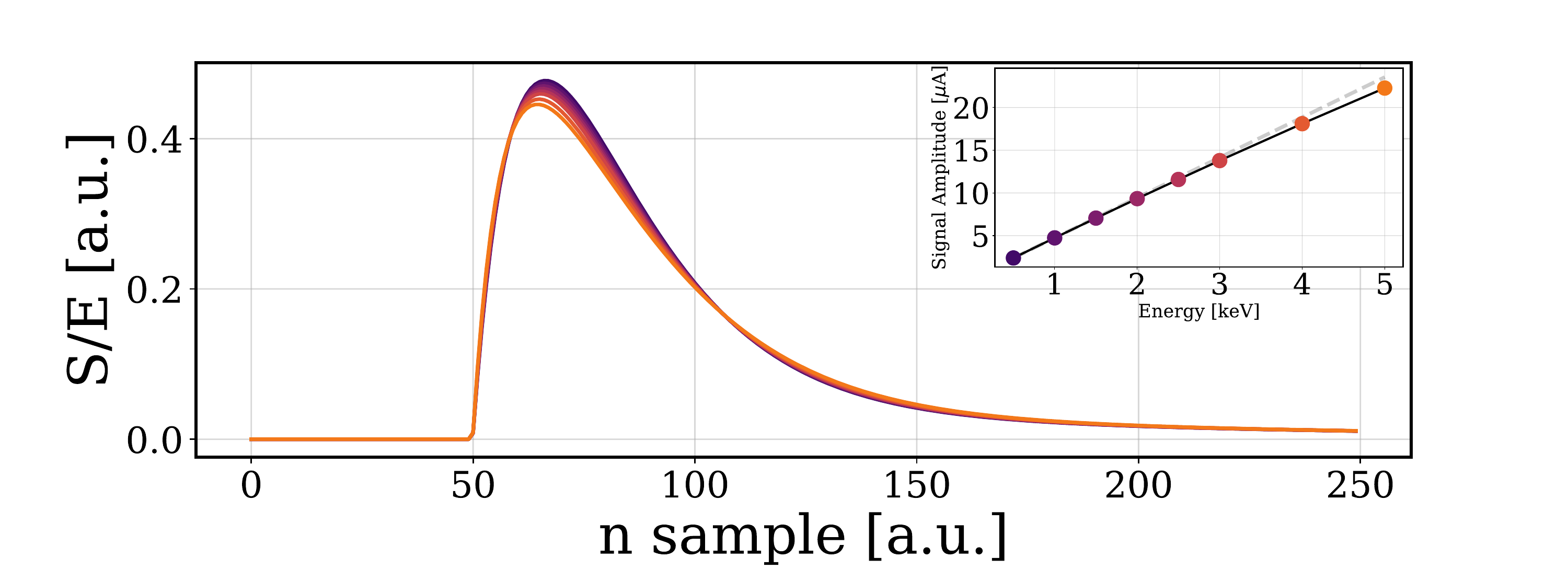}%
}
	\caption{Pulse profiles corresponding to different energies from 0.5 to 5 keV for two different detectors with non-linear (det. [a]), and nearly-linear response (det. [b]).	
	In order to compare the signals, all the pulses are normalized by dividing their amplitude by the energy.
}
	\label{fig:signal}       % Give a unique label
\end{figure}

\noindent To test the DSVP effectiveness with slightly different signal shapes, the physical parameters in the differential equations are chosen to describe three types of detectors (Figure \ref{fig:signal}):
\begin{enumerate}
    \item [a.] the detectors in \cite{Alpert:2015mtu} which are characterized by a non linear response and one thermal body.
    \item[b.] the target detectors of HOLMES \cite{alpert2019high} have nearly-linear response and behave according to a two thermal body model.
    \item[c.] same nominal design as b. except for the production process, causing a significantly weaker link toward the thermal bath. Thus, the signals have a slower decay time and a lower signal amplitude respect to b. . Despite this difference, the detectors show a linear response to energy deposition with a two-body feature.

\end{enumerate}
 
\noindent The detectors of type b. are the most promising for the HOLMES goals. Therefore, for this detector three different configurations were tested, in which the rise time was changed, adjusting the inductance of the circuit and keeping the other detector parameters constant.

\subsection{DSVP \& HOLMES}
\label{sec:3c}
We indicate the ratio between the number of  pile-up pulses and the number of single pulses in the ROI as $f_{pp}^{ROI}$. From simulations, setting a time resolution of 10 $\mu$s a value of $f_{pp}^{ROI} \simeq 2$ is expected. 
\noindent In order to apply the algorithm, the $\vec{M}$ matrix, which contains the ROI events, must have $N_A > N_B$, thus  $f_{pp}^{ROI}$ needs to be lowered below one.
\noindent To reduce this ratio many different strategies can be adopted. In the following a non exhaustive list is reported.

\begin{itemize}
	\item \textit{Adding an additional calibration source.} By adding a source characterized by a monochromatic X-ray emission in the ROI, the number of single pulses in the ROI can be increased while keeping the number of pile-up pulses unchanged. This approach can be very useful because it reshapes the energy spectrum, potentially reducing the probability of discarting single events with energy very close to the end-point.
	A similar approach was investigated by Alpert  \cite{Alpert:2015mtu}. %We did not explore this strategy in our simulations, because we want to see what results we can achieve using only the events coming from the $^{163}$Ho.
	
	\item \textit{Volumetric cuts.} The events of the training dataset $\vec{T}$ are distributed in a finite volume in the $k$-dimensional projection space. The single pulses in the ROI reasonably lie within the same portion of space, while the pile-up are expected to be distributed in a different region.	 
	Thus if we select only the points in the projection space lying inside the volume which includes the training dataset, we could easily eliminate a large fraction of pile-up events.
	\noindent %We choose as training dataset the events at the M1 peak of $^{163}$Ho spectrum. This region satisfy the condition of $N_B << N_A$ because the $f_{pp}$ is expected to be lower than [].   
	Before evaluating their projection on the $\vec{T}$ right singular vectors, the $\vec{T}$ and ROI events are normalized to set their amplitude equal to one. 
	Then, we define the region in the $k$-space in which the $\vec{T}$ events are distributed. We increase it by a little amount in order to account small non-linearity effects. Finally, we select only the events in the ROI included inside this region.
	This method can achieve good time resolution, but it works only if the detector response does not depart from linearity too much, so in our simulation in detectors $b.$ ,  $c.$  but not $a.$ .
	
	\item \textit{Filtering.} Few filtering techniques allow to achieve effective time resolution close to the sampling time. Among these, a particular Wiener filter, as described in \cite{Ferri:2016ajd}, is probably the best technique to achieve this goal.
	
\end{itemize}

\begin{figure}[h!]
	\centering
	\includegraphics[width=1.\linewidth]{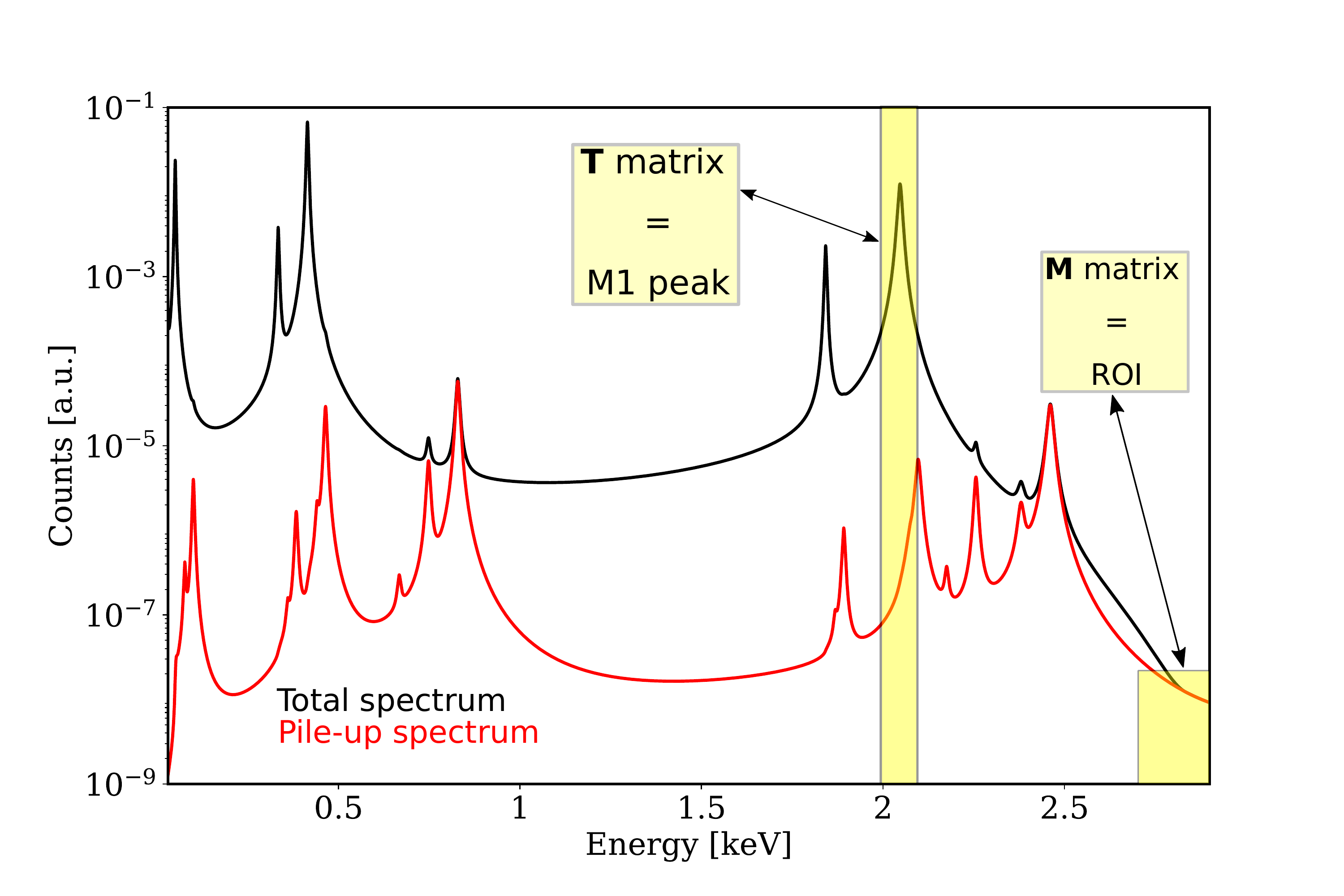}
	\caption{De-excitation simulated spectra of $^{163}$Ho with $f_{pp}^{tot} = 3\times10^{-3}$. Near the end-point the single pulse counts are outnumbered by pile-up counts.}
	\label{fig:roispectrum}
\end{figure}

\noindent For the HOLMES purposes, in order to fulfill the $N_A > N_B$ condition in the ROI the most suitable and practical method are the 'wiener filter' and the 'volumetric cuts'. As indicated in Table \ref{tab:2}, applying these algorithms to the ROI events, the $f_{pp}^{ROI}$ can be reduced around $0.6$. 
Most of our simulations are therefore aimed to test their applications. Nevertheless, in sec \ref{sec:4b} the performance of the DSVP technique with an external calibration source is also shown.

\noindent  As shown in Fig \ref{fig:roispectrum}, we use the events at M1 peak ($E \sim 2$ keV) as training region $\vec{T}$.
This is also the energy range in which the average signal for the Wiener filter is defined.
The M1 peak is the most suitable region for two reasons: it is the peak closest to the ROI, thus reducing the non-linearity effects on the filters and on the discrimination algorithm and it fulfills the condition of $N_B << N_A$. The $f_{pp}$ in this sector is expected to be of the order of $\sim 10^{-3}$, which can be further reduced with a \textit{raw cleaning with PCA}, as described in section \ref{sec:2a}.

%For each detector type we have adopted the most suitable method between the 'wiener filter' and the 'volumetric cuts' in order to fulfill the $N_A > N_B$ requirement in the ROI. As indicated in table [], using these the initial $f_{pp}^{ROI}$ is reduced around $0.6$, fulfilling the initial condition of the DSVP algorithm for the matrix $\vec{M}$ ($N_B \simeq 0.6 \times N_A$).

\section{Simulation results}
\label{sec:4}
\noindent To quantify the efficiency of the pile-up discrimination algorithms, we define an effective time resolution $\tau_{eff}$ as the ratio of the number of retained piled-up records to single-pulse records after the algorithm divided by the same ratio referred to raw data, times 10 $\mu$s.
\begin{equation}
\label{eq_tau_eff}
\tau_{eff} = \biggl(\frac{ pup}{ single}\biggr)_{final} \div \biggl(\frac{ pup}{ single}\biggr)_{initial} \times 10 \ \mu s
\end{equation}

\noindent Through simulations similar to the ones in \cite{nucciotti2014statistical}, \cite{nucciotti2010expectations}, we have preliminary estimated that even a small fraction of false negative modifies the single events spectrum and leads to a systematic error on the neutrino mass evaluation. We note that in our simulations no single pulse event was mistaken as pile-up. The DSVP technique described in section \ref{sec:2} is designed to leave unaffected the $A$ events.

\noindent In applications where a more robust discrimination of the $B$ events is required, it is possible to adapt the algorithm toward this goal, for example by adjusting the threshold definition (Eq. \ref{thr_def}), at the expenses of increasing the chance of deleting some $A$ events.

\noindent The energy dependence of the method must be assessed for each specific application. In general, we would say that the events with the energies further away than the mean energy in the dataset $\vec{M}$ are most likely to become false positive.
In the end, the number of false positives is due to the threshold value (Eq. \ref{thr_def}), while their nature is due to the mean “morphology” of the events present in the original dataset $\vec{M}$.

\subsection{DSVP with Wiener Filter and Volumetric cuts}
\begin{table*}
	\caption{Effective time resolution of the DSVP with Wiener Filter and Volumetric cuts. We indicated with (*) the algorithm used in that simulation to lower the ratio of $f_{pp}^{ROI}$ below one. For simplicity, we always set $N_B$ equal to the exact number of pile-up pulses in the ROI. The errors associated with the DSVP $\tau_{eff}$ are $ \leq 5\%$ and are due to the random nature of the modified minimization RANSAC algorithm.}
	\label{tab:2}       % Give a unique label
	\begin{tabular}{cccccc}
		\hline\noalign{\smallskip}
		Detector type & Rise Time [$\mu$s] & $t_{sample}$ [$\mu$s] &  $\tau_{eff}$ Wiener Filter &  $\tau_{eff}$ Volumetric cuts & $\tau_{eff}$ with DSVP   \\
		\noalign{\smallskip}\hline\noalign{\smallskip}
		b. & 11 & 2  & 2.26 & 2.12 (*) & 1.55 \\
		
		b. & 17 & 2 & 2.37 (*) & 2.60 & 1.55 \\
		
		b. & 22 & 2 & 2.94  & 2.90 (*) & 2.01 \\
		
		b. & 17 & 1 & 1.66 (*) & 2.00 & 0.94 \\
		
		a. & 10 & 2 & 1.82 (*) & - & 1.24 \\
		
		c. & 19 & 2 & 2.70 (*) & 3.54 & 1.82 \\
		
		\noalign{\smallskip}\hline
	\end{tabular}
\end{table*}

\noindent We have estimated the $\tau_{eff}$ on the simulated data processed with the DSVP after lowering the initial $f_{pp}^{ROI}$ using the 'wiener filter' or the volumetric cut techniques. 

\noindent Furthermore, before being processed by the DSVP algorithm, the signals where also whitened, i.e. transformed to whiten noise by a fast Cholesky-factor backsolve procedure \cite{fowler2015microcalorimeter}.
\noindent The results are reported in Table \ref{tab:2}. All the simulations showed that the DSVP is able to reach a time resolution lower than the sampling time of the signal.

%\noindent We also tested the dependence of the algorithm on the sampling time and the rise time of the pulse.
\noindent Table \ref{tab:2} shows that the time resolution is strongly dependent on the sampling time, the faster the better, but also on the rise time of the pulse. While the sampling frequency is constrained by the readout resources, there is more scope to change the rise time of the detectors, acting on the electrical time constant of the biasing circuit. Changing the rise time by a factor two may be achieved reducing the inductance of the TES circuit by a similar factor. This change the noise spectrum too but usually it does not worsen the energy resolution.

\noindent Also, the non-linear detector response generally improves the efficiency of pile-up recognition algorithms. 
When two near-coincident energy depositions happen inside the TES, the detector will have different starting conditions.
The shape of the pile-up pulse will be much more different from the single pulse for a non-linear TES than for a linear one, thus allowing the algorithms to recognize them more efficiently.

\noindent As we stressed in section \ref{sec:2}, the only external parameter required by the DSVP algorithm is the number of events that it should discard at most, $N_B$.
To quantify the influence of this parameter on the effectiveness of the algorithm, we fixed the dataset $\vec{M}$ and varied $N_B$, computing the effective time resolution each time. 

\begin{figure*}[h!]
	\centering
	\includegraphics[width=1.\linewidth]{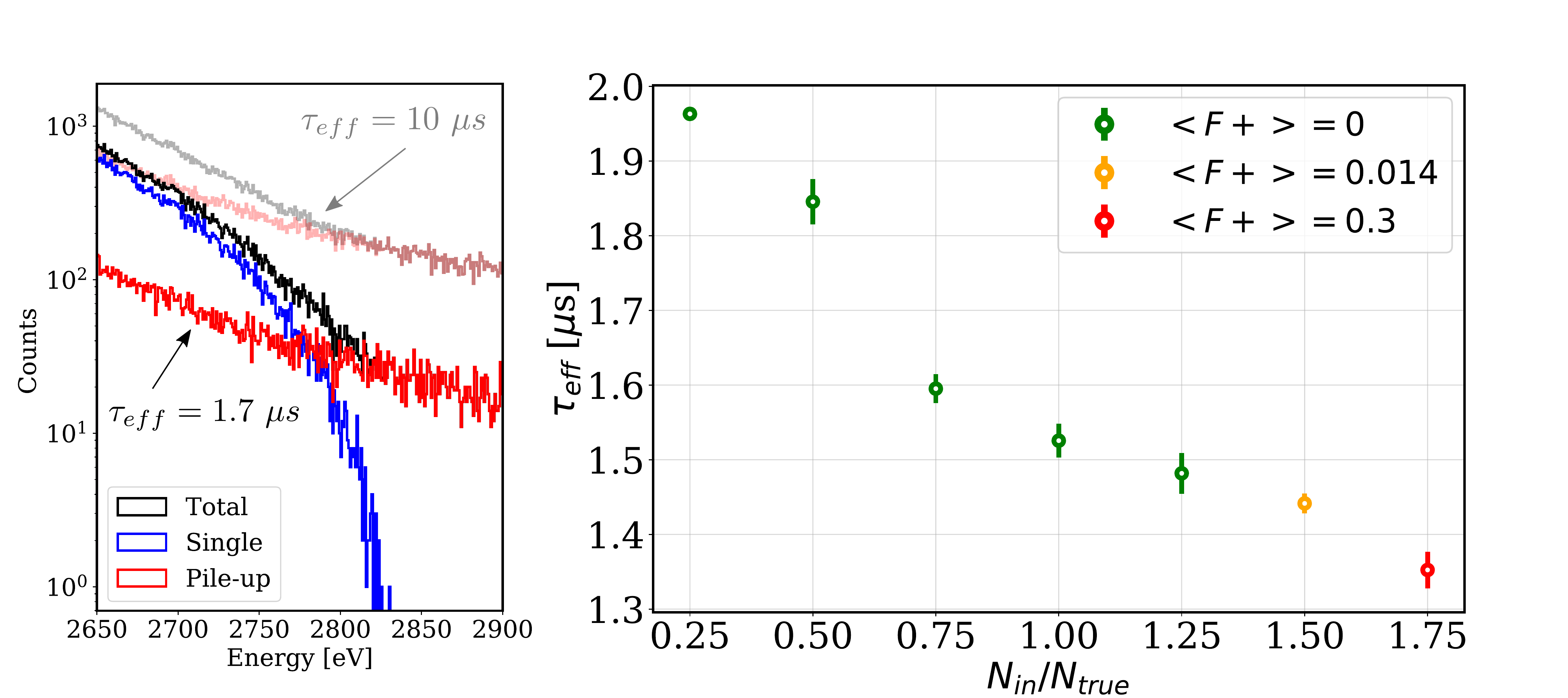}
	\caption{\textit{Left panel}: ROI energy spectrum before and after the application of the WF/Volumetric cuts and the DSVP technique. Light line represents the energy spectrum with $\tau_{eff}$ of 10 $\mu$s, while the solid line with a $\tau_{eff}$ of 1.7 $\mu$s. \textit{Right panel}: The dependence of $\tau_{eff}$ and the average percentage of false positive $F+$ from the input parameter $N_B$ ($N_{in}$), which is normalized by the number of pile-up pulses present in the ROI ($N_{true}$).}
	\label{fig:tau_eff_N}
\end{figure*}

\noindent Figure \ref{fig:tau_eff_N} shows that no false positive was detected even if we get the number of event to eliminate wrong up to 50\%. 

\subsection{DSVP with additional calibration source}
\label{sec:4b}
\begin{figure*}[h!]
	\centering
	\includegraphics[width=1.\linewidth]{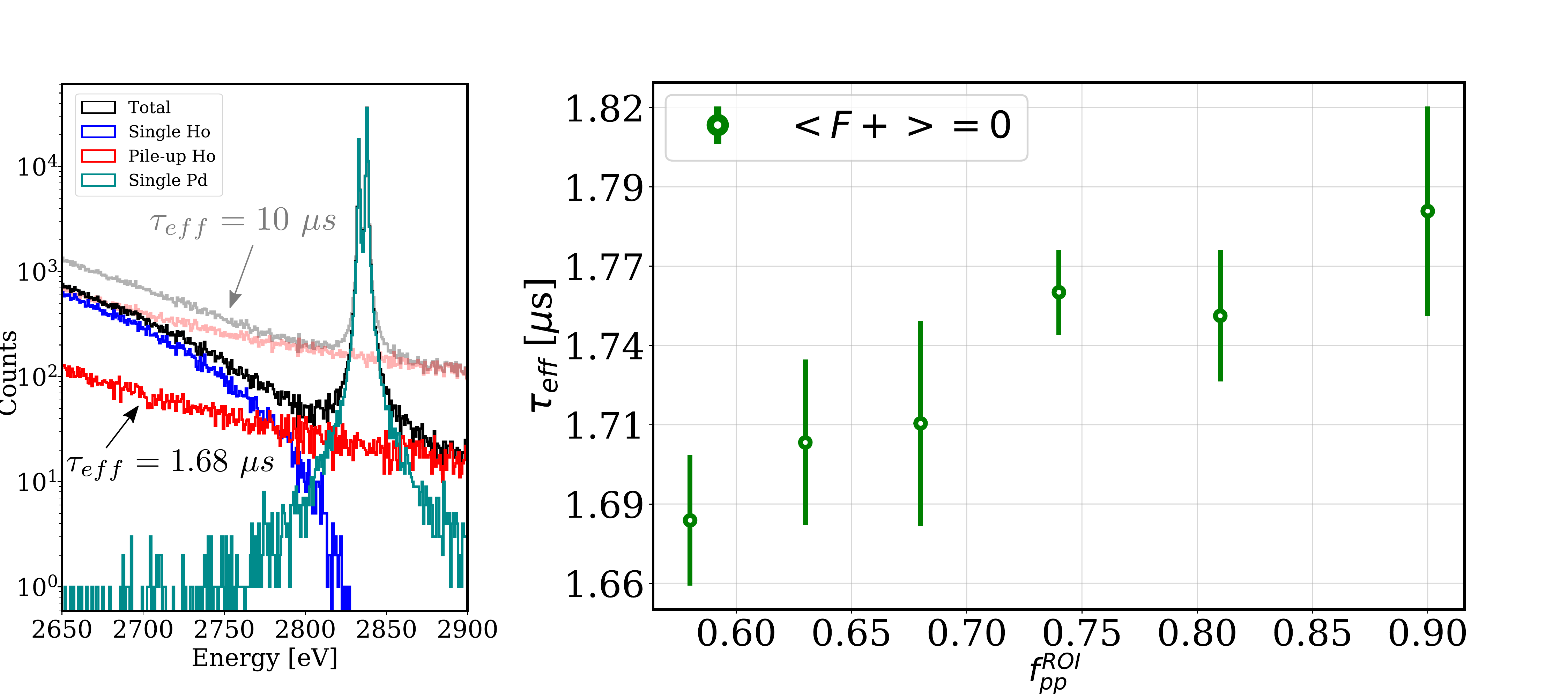}
	\caption{\textit{Left panel}: ROI energy spectrum with Pd L$\alpha$ peaks before and after the application of the DSVP technique. The initial $f_{pp}^{ROI}$ was set to 0.58. \textit{Right panel}: The dependence of $\tau_{eff}$ from the $f_{pp}^{ROI}$ is shown for the detector $b.$ with a rise time of 17 $\mu$s and a sampling time of 2 $\mu$s. In this case, $N_B$ was equal to the number of pile-up pulses in the ROI.}
	\label{fig:tau_eff_Pd}
\end{figure*}

We have also tested if the performance of the DSVP technique remains unchanged reducing the $f_{pp}^{ROI}$ by adding an external source of single events with energy inside the region of interest instead of using preliminary filters. 
We added a source from L$\alpha$ x-ray emission lines of Pd (2.833, 2.839 keV). %[We slightly modified the DSVP algorithm in order to restrict the model curves inside the region of the $\vec{M'}$ points. METTO QUESTA FRASE?]. 
Figure \ref{fig:tau_eff_Pd} shows that increasing the number of photons of the Pd source (thus decreasing $f_{pp}^{ROI}$) the effective time resolution of the DSVP improves. Moreover, $\tau_{eff}$ always remains below the sampling time even for a pile-up fraction up to 0.9.

%with our current experiment design but without the fine pile-up recognition algorithms described in this article, we need to broaden the measurement time by roughly a factor of 4: from 3 to 12 years.

\section{Conclusions}

\noindent The DSVP algorithm represents a very powerful technique to decrease the number of undesirable events in a dataset.

\noindent In this work we have applied this algorithm for pile-up discrimination, which can lead to major improvement in experimental sensitivity for experiments such as HOLMES (neutrino mass measurement) or CUPID ($0\nu\beta\beta$) \cite{Chernyak:2012zz}.
It can also be useful to recognize single-site events of the $0\nu\beta\beta$ interactions from multi-site background events in GERDA \cite{Agostini:2013jta}.
We tested the DSVP technique for the HOLMES application and we compared its efficiency, represented in this case by the effective time resolution $\tau_{eff}$, to more 'classical' discrimination techniques, resulting in a better time resolution.

\noindent With the target detector of HOLMES, the DSVP techniques allows us to reduce the total fraction of pile-up events from $10^{-3}$ ($\sim \tau_{eff} \ 3 \ \mu s$) to $10^{-4}$ ($\sim \tau_{eff} \ 1.5 \ \mu s$), thus improving the neutrino mass sensitivity from 2 eV to about 1.4 eV.
To put this result in perspective, achieving the same improvement would require to increase the acquisition time by a factor 4: from 3 to 12 years.

%\begin{acknowledgements}

%\end{acknowledgements}

\bibliographystyle{spphys}       % APS-like style for physics
\bibliography{pup-holmes-epjc}   % name your BibTeX data base

\end{document}